\documentstyle[12pt]{article}
\title{Quasi-set-theoretical foundations of statistical mechanics: a research program}
\author{{\bf Adonai S. Sant'Anna}\\{\small \it Dep. Matem\'atica, UFPR, C. P. 019081}\\{\small \it Curitiba, PR, 81531-990, Brazil} \and {\bf Alexandre M. S. Santos}\\{\small \it Dep. F\'{\i}sica, UFPR, C. P. 019081}\\{\small \it Curitiba, PR, 81531-990, Brazil}}
\date{ }
\begin{document}

\newtheorem{definicao}{\bf Definition}
\newtheorem{teorema}{\bf Theorem}
\newtheorem{lema}{\bf Lemma}
\setlength{\unitlength}{1mm}
\renewcommand{\thefootnote}{\arabic{footnote}}
\setcounter{footnote}{0}
\newcounter{cms}
\setlength{\unitlength}{1mm}

\maketitle

\begin{abstract}
Quasi-set theory provides us a mathematical background for dealing with collections of indistinguishable elementary particles. In this paper, we show how to obtain the usual statistics (Maxwell-Boltzmann, Bose-Einstein, and Fermi-Dirac) into the scope of quasi-set theory. We also show that, in order to derive Maxwell-Boltzmann statistics, it is not necessary to assume that the particles are distinguishable. In other words, Maxwell-Boltzmann statistics is possible even in an ensamble of indistinguishable particles, at least from the theoretical point of view. The main goal of this paper is to provide the mathematical grounds of a quasi-set-theoretical framework for statistical mechanics.
\end{abstract}

\section{Introduction}

	According to any textbook about statistical mechanics, we know that Maxwell-Boltzmann (MB) statistics gives us the most probable distribution of $N$ {\em distinguishable\/} objects into, say, boxes with a specified number of objects in each box. In this paper we show that the hypothesis concerning distinguishable objects is unnecessary. Usually, classical and quantum distribution functions are mathematically derived in a naive fashion; but in our case an axiomatic framework is necessary if we want to show that individuality is not a necessary assumption in classical statistical mechanics. Classical logic and mathematics are commited with a conception of identity which does not make any distinction between identity and indistinguishability: indistinguishable things are the very same thing and conversely. In several standard textbooks on quantum mechanics, for example, there is no clear distinction between indistinguishability and identity. So, it is necessary to settle some philosophical terms in order to avoid confusions. When we say that $a$ and $b$ are {\em identicals\/}, we mean that they are the very {\em same\/} individual, that is, there are no `two' individuals at all, but only one which can be named indifferently by either $a$ or $b$. By {\em indistinguishability\/} we simply mean agreement with respect to attributes. We recognize that this is not a rigorous definition. Nevertheless such an intuition is better clarified in the next Sections.

	Our proposed axiomatic framework for dealing with quantum and classical statistics is quasi-set theory \cite{Krause-92} \cite{Krause-96}. Quasi-set theory ${\cal Q}$ allows us the presence of two sorts of atoms (atoms in the mathematical sense, that is, {\it Urelemente\/}), termed $m$-atoms and $M$-atoms, identified by two unary predicates $m(x)$ and $M(x)$, respectively. Concerning the $m$-atoms, a weaker `relation of indistinguishability' (denoted by the symbol $\equiv$), is used instead of identity, and it is postulated that $\equiv$ has the properties of an equivalence relation. The predicate of equality cannot be applied to the $m$-atoms, since no expression of the form $x = y$ is a well formed formula if $x$ or $y$ denote $m$-atoms. Hence, there is a precise sense in saying that $m$-atoms can be indistinguishable without being identical. In standard mathematics, when we say that $x = y$ ($x$ is identical to $y$) we are talking about the very same object, with two different names: $x$ and $y$. The axioms of quasi-set theory are a very natural extension of the axioms of Zermelo-Fraenkel (ZF) set theory.

	This work is part of a research program concerning the problems of non-individuality in quantum mechanics and related topics. In previous works it was presented a manner to cope with collections of `physically' indistinguishable particles in a {\em set-theoretical framework\/} (that is, standard set theory) by using hidden variables \cite{Sant'Anna-97a} \cite{Sant'Anna-97b}. Quasi-set theory was first proposed as a mathematical framework for quantum distributions in \cite{Krause-99}, where a quasi-set-theoretical predicate for collections of indiscernibles was presented and Bose-Einstein (BE) and Fermi-Dirac (FD) statistics were derived. Here we show a simpler manner to derive quantum statistics based on quasi-set theory and we also discuss about MB statistics. We propose a generalization ${\cal Q'}$ of quasi-set theory and show that MB statistics is possible even in an ensamble of indistinguishable particles.

	By using quasi-set theory instead of standard set theory, our paper provides a way of obtaining the statistics from the assumption that the `non-individuality' of quantum objects should be ascribed right at the start. In the set-theoretical picture presented in \cite{Sant'Anna-97a} and \cite{Sant'Anna-97b}, such an assumption is not (and cannot be) stated.

	In the next Section we present quasi-set theory. In Section 3 we show how to derive MB statistics in a collection of indistinguishable objects. In Section 4 we make a generalization of quasi-set theory, which we call ${\cal Q'}$, and show that quantum statistics may be seen as a special case of MB statistics. In Section 5 we prove that ${\cal Q'}$ is consistent iff ZF set theory is consistent. In Section 6 we present the main conclusions and make a conjecture in the context of this quasi-set theoretical approach to statistical mechanics.

\section{Quasi-set theory}

\subsection{The Language}

	This Section is essentially based on \cite{Krause-96}, which is variant of the formulation presented in \cite{Krause-92}.

	The language of quasi-set theory ${\cal Q}$ is that of the first order predicate calculus {\it without\/} identity.  The intuitive idea is to allow the existence of {\it Urelemente\/} of two kinds, which are called $m$-atoms and $M$-atoms. The latter act as atoms of ZFU (Zermelo-Fraenkel with {\it Urelemente\/}), while the former are supposed to be objects to which the concept of identity cannot be applied in a sense.

	The specific symbols of ${\cal Q}$ are three unary predicates $m$, $M$ and $Z$, two binary predicates $\equiv$ and $\in$ and an unary functional symbol $qc$. Terms and well-formed formulas are defined in the standard way, as are the concepts of free and bound variables, etc.. We use $x$, $y$, $z$, $u$, $v$, $w$ and $t$ to denote individual variables, which range over quasi-sets (henceforth, qsets) and {\it Urelemente\/}.  Intuitively, $m(x)$ says that `$x$ is a microobject' ($m$-atom), $M(x)$ says that `$x$ is a macroobject' ($M$-atom) while $Z(x)$ says that `$x$ is a set'. The term $qc(x)$ stands for `the quasi-cardinal of (the qset) $x$'. The {\it sets\/} are supposed to be exact copies of the sets in ZFU. 

	The formulas  $\forall_{P} x (\ldots)$ and $\exists_{P} x (\ldots)$ abbreviate $\forall x (P(x) \to (\ldots))$ and  $\exists x ( P(x) \wedge (\ldots)$ respectively, where $P$ is a predicate of the language, $\to$ is the conditional of propositional calculus, and $\forall$ and $\exists$ are, respectively, the universal and the existential quantifiers of predicate calculus. We use further standard logical notation: $\neg$ is negation, $\wedge$ is conjunction, $\vee$ is disjunction, and $\leftrightarrow$ is biconditional.

\begin{definicao}
\begin{enumerate}
\item $Q(x)  := \neg (m(x) \vee M(x))$  ($x$ is a {\em quasi-set})
\item $P(x)  := Q(x) \wedge  
\forall y (y \in x \to m(y))$ 
 ($x$ is a {\em `pure' quasi-set}, that is, a quasi-set whose elements are
$m$-atoms only).
\item $D(x)  := M(x) \vee Z(x)$  ($x$ is a  {\em classical object\/},
or {\em Dinge\/}, in Zermelo's original sense, that is, $x$ is a (classical) 
{\it Urelement\/} or a set). 
\item $E(x)  := Q(x) \wedge \forall y (y \in x \to Q(y))$
\item {\rm [{\em  Extensional Equality\/}]}  For all  $x$ and $y$, 
if they are not  $m$-atoms, then:
$$x =_{E} y := \forall z ( z \in x \leftrightarrow z \in y ) \vee
(M(x) \wedge M(y) \wedge x \equiv y)$$
\item {\rm [{\em Subquasi-set]}}  For all $x$ and $y$,
if they are not atoms, then:
 $$x \subseteq y := \forall z (z \in x \to z \in y)$$
\end{enumerate}
\end{definicao}

\noindent
If $x \neq_{E} y$, that is, $\neg (x =_{E} y)$, we say that $x$ and $y$ are {\em extensionally distinct\/}. As is usual, $x \subset y$, means  $x \subseteq y \wedge x \neq_{E} y$. It is immediate that $x \subseteq y \wedge y \subseteq x \to x =_{E} y$.

\subsection{The First Axioms}

	The first four axioms of ${\cal Q}$ are {\it The Axioms of Indistinguishability\/}:

\begin{description}

\item[Q1] $\forall x (x \equiv x)$

\item[Q2] $\forall x \forall y (x \equiv y \to y \equiv x)$

\item[Q3] $\forall x \forall y \forall z (x \equiv y \wedge y \equiv z \to x \equiv z)$

\item[Q4] $\forall x \forall y ( D(x) \wedge D(y)  \to (x \equiv y \to (A(x,x) \to A(x,y))))$, with the usual syntactic restrictions.

\end{description}

	Axiom {\bf Q4} excludes $m$-atoms from the substitutivity law since if substitutivity is postulated to include them as well, then {\bf Q1}--{\bf Q4} turn to be exactly the axioms usually used for the predicate of identity \cite{Mendelson-97} and no syntactical difference between identity and indistinguishability could be achieved.  By using {\bf Q4} as above, we preserve Leibniz Law of Identity of Indiscernibles for the `macroscopic' (that it, those which are not $m$-atoms) indistinguishable objects (including qsets).

\begin{description}

\item[Q5] No {\it Urelemente\/} is at the same time an $m$-atom and an $M$-atom:
$$\forall x ( m(x) \vee M(x) \to \neg (m(x) \wedge M(x)))$$

\item[Q6] If $x$ has an element, then $x$ is a qset. In other words, the atoms are empty:
$$\forall x \forall y (x \in y \to Q(y))$$

\item[Q7] Every set is a qset:
$$\forall x (Z(x) \to Q(x))$$

\item[Q8] No  set contains $m$-atoms as elements:
$$\forall x (\exists_{m} y (y \in x) \to \neg Z(x))$$

\item[Q9] Qsets whose elements are `classical objects' are sets and conversely:
$$\forall_{Q} x (\forall y (y \in x \to D(y)) \leftrightarrow Z(x))$$

\end{description}

\begin{teorema}
If $x$ is an $M$-atom (respectively, a qset) and $x \equiv y$, then $y$ is also an $M$-atom (respect., a qset).
\end{teorema}

\noindent
{\bf Proof:\/} (See \cite{Krause-96})

	If $x$ is an $m$-atom, the analogous case of the above theorem cannot be proven. For details see \cite{Krause-96}. So, we need the following postulate:

\begin{description}

\item[Q10] Objects which are indistinguishable from $m$-atoms are also $m$-atoms:
$$\forall x (m(x) \wedge x \equiv  y \to m(y))$$

\end{description}

	From the above axioms, it follows that {\it sets\/} cannot have $m$-atoms as elements and, in order its elements also be `classical', they also cannot have $m$-atoms as elements, and so on. Hence this idea pervades the `interior' of the elements of a qset, and this implies that a qset is a set iff its transitive closure (this concept can be defined in the standard way) does not contain $m$-atoms.

\begin{description}

\item[Q11] There exists a qset (denoted `$\emptyset$') which is a set and which does not have elements: 
$$\exists_{Z} x \forall y (\neg (y \in x))$$ 

\end{description}

\begin{definicao}{\em [Similar quasi-sets]} 
 For all quasi-sets  $x$ and $y$,  
$$Sim(x,y) := \forall z \forall t (z \in x \wedge t \in y \to z \equiv t)$$
\end{definicao}

	Intuitively, similar qsets have as elements objects `of the same sort'.  The idea of `objects of the same sort' can be realized by passing the quotient by the relation of indistinguishability. This procedure defines equivalence classes of indistinguishable objects and, if they are `classical', the classes turn to be unitary sets, since the indistinguishability relation coincides with equality in this case

\begin{description}

\item[Q12] Indistinguishable sets are extensionally identicals:
$$\forall_{Z} x \forall_{Z} y (x \equiv y \to x =_{E} y)$$

\end{description}

	{\bf Q12} imposes the requeriment that the usual extensional properties of the sets of ZFU are valid for the sets of ${\cal Q}$. Further explanations regarding this axiom are presented after the axiom {\bf Q27}.

\begin{description}

\item[Q13] [{\it `Weak-Pair'\/}] For all  $x$ e $y$, there exists a qset whose elements are the indistinguishable objects from either $x$ or $y$:
$$\forall x \forall y \exists_{Q} z \forall t (t \in z \leftrightarrow t \equiv x \vee t \equiv y)$$

\end{description}

	The weak-pair of $x$ and $y$ is denoted $[x, y]$ and in the case when $x$ and $y$ are both classical objects, we may use the standard notation $\{x, y \}$, since in this case the only things indistinguishable from $x$ and $y$ will be respectively $x$ and $y$ themselves. If $x \equiv y$, we denote the weak-pair by $[x]$, called the {\it weak-singleton\/} of $x$, which is the qset of that which is indistinguishable from $x$.  It is important to realize, as it will be clear below, that it is consistent with the theory to admit that the weak-singleton of $x$ may have quasi-cardinal greater than one. In this sense, ${\cal Q}$ {\it allows\/} the existence of indistinguishable objects which cannot be said to be identical.

\begin{description}

\item[Q14] [{\em  The Separation Schema\/}] By considering the usual syntactical restrictions on the formula $A(t)$, we have: 
$$\forall_{Q} x \exists_{Q} y \forall t (t \in y \leftrightarrow t \in x \wedge A(t))$$

\end{description}

	This qset will be written $[t \in x : A(t)]$. The separation axiom allows us to form subquasi-sets of a quasi-set $x$ by considering those elements of $x$ that satisfy a certain property expressed (in the language of ${\cal Q}$)  by a formula $A(t)$.  This idea conforms itself with the  intended interpretation of the $m$-atoms as elementary particles, since in ordinary physics it is possible to `select', from a certain collection of elementary particles, a certain number of them that satisfy a particular condition.

\begin{description}

\item[Q15] [{\em Union\/}] $\forall_{Q} x (E(x) \to \exists_{Q} y (\forall z (z \in y) \leftrightarrow \exists t (z \in t \wedge t \in x)))$

\end{description}

	As usual, this qset is written $$\bigcup_{t \in x} t$$ and we  still write $x \cup y$ in the same sense as in the standard set theories. 

\begin{description}

\item[Q16] [{\em Power-qset}] $\forall_{Q} x \exists_{Q} y \forall t (t\in y \leftrightarrow t \subseteq x)$ 

\end{description}

	The power qset of $x$ is denoted by ${\cal P}(x)$.

\begin{definicao}
\begin{enumerate}
\item  $\overline{x} := [y \in x : m(y)]$ 
\item  $\langle x, y \rangle := [[x],[x,y]]$ (the generalized ordered pair)
\item  {\rm For every quasi-sets $x$ and $y$,} $ x \times y := [\langle z, u \rangle \in {\cal P}{\cal P}(x \cup y) : z \in x \wedge u \in y]$
\item The intersection $x \cap y$ of two quasi-sets can be defined do that $t \in x \cap y$ iff $ t \in x \wedge t \in y$ as usual.
\end{enumerate}
\end{definicao}

\begin{description}
\item[Q17] [{\em Infinity}] $\exists_{Q} x (\emptyset \in x \wedge \forall y (y \in x \wedge Q(y) \to y \cup [y] \in x))$

\item[Q18] [{\em Regularity\/}] Quasi-sets are well-founded, that is, for every qset $x$, there are no infinite chains $\ldots \in x_{2} \in x_{1} \in x$:
$$\forall_{Q} x (E(x) \wedge x \neq \emptyset \to \exists_{Q} (y \in x \wedge y \cap x = \emptyset))$$

\end{description}

\subsection{Quasi-Relations} 

	The concept of {\em relation\/} and in particular that of {\em equivalence relation\/} is like the standard one: $w$ is a relation between two quasi-sets $x$ and $y$ if $w$ satisfies the following predicate $R$: 

$$R(w) := Q(w) \wedge \forall z (z \in w \to \exists u \exists v (u \in x \wedge v \in y \wedge z =_{E} \langle u, v \rangle))$$ 

	As in the classical case, $R \in {\cal P}{\cal P}{\cal P}(x \cup y)$. Furthermore, as usual, if $x =_{E} y$, we say that $R$ is a relation {\em on\/} $x$. We denote by $Dom(R)$ (the {\em domain} of $R$) the qset $[u \in x : \langle u, v \rangle \in R ]$ and by $Rang(R)$ (the {\em range} of $R$) the qset $[v \in y : \langle u, v \rangle \in R]$.  

	A particular interesting case of an equivalence relation on a qset $x$ is the indistinguishability relation, which satisfies the predicate $R$ above and, due to the axioms {\bf Q1} -- {\bf Q3}, has the required properties. In this case, if $x$ is a pure qset, then the `quotient qset' $x/ \equiv$ stands for a collection of equivalence classes of indistinguishable objects.

\begin{teorema}
No partial, total or strict order relation can be defined on a pure qset whose elements are  indistinguishable  from one another.
\end{teorema}

\noindent
{\bf Proof\/:} (see \cite{Krause-96}).

\subsection{Axioms of Quasi-Cardinals}

\begin{description}

\item [Q19] Every object which is not a qset (that is, every  {\it Urelement\/}) has quasi-cardinal zero:
$$\forall x (\neg Q(x) \to qc(x) =_{E} 0)$$

\item[Q20] Every qset has an unique quasi-cardinal which is a cardinal (as defined in the `copy' of ZFU) and, if the qset is in particular a set, then this quasi-cardinal is its cardinal stricto sensu:

$$\forall_{Q} x \exists ! y (Cd(y) \wedge y =_{E} qc(x) \wedge (Z(x) \to y =_{E} card(x)))$$

\item[Q21] Every non-empty qset has a non null quasi-cardinal:
$$\forall_{Q} x (x \neq_{E} \emptyset \to qc(x) \neq_{E} 0)$$

\end{description}

	The next axiom says that if the quasi-cardinal of a qset $x$ is $\alpha$, then for every quasi-cardinal $\beta \leq \alpha$, there is a  a subquasi-set of $x$ whose quasi-cardinal is $\beta$.

\begin{description}

\item[Q22] $\forall_{Q} x (qc(x) =_{E} \alpha \to \forall \beta (\beta \leq_{E} \alpha \to \exists_{Q} y (y \subseteq x \wedge qc(y) =_{E} \beta))$

\item[Q23] The quasi-cardinal of a subquasi-set of $x$ is not greater than the quasi-cardinal of $x$:
 $$\forall_{Q} x \forall_{Q} y (y \subseteq x \rightarrow qc(y) \leq_{E} qc(x))$$

\item[Q24] $\forall_{Q} x \forall_{Q} y (Fin(x) \wedge x \subset y \to qc(x) < qc(y))$, where $Fin(x)$ corresponds to say that $x$ is finite.

\item[Q25] $\forall_{Q} x \forall_{Q} y (\forall w \neg (w \in x \wedge  w \in y) \to qc(x \cup y) =_{E} qc(x) + qc(y))$

\end{description}

	In the next axiom,  $2^{qc(x)}$  denotes (intuitively) the quantity of subquasi-sets of $x$. Then, 

\begin{description}

\item[Q26] $\forall_{Q} x (qc({\cal P}(x)) =_{E} 2^{qc(x)})$

\end{description}

	This last axiom is one of our central interests in this paper, as we see below.

\subsection{`Weak' Extensionality}

	We begin by recalling that the quasi-sets $x$ and $y$ are {\em similar\/}, ($Sim(x,y)$) -- cf. Definition (2) -- if their elements are indistinguishable. Then, we define: 

\begin{definicao}
The quasi-sets $x$ and $y$ are {\rm Q-Similar} if they are similar and have the same quasi-cardinality.
\end{definicao}

	By observing that the quotient quasi-set $x/ \equiv$ may be regarded as a collection of equivalence classes of indistinguishable objects, the weak axiom of extensionality is stated as:

\begin{description}

\item[Q27] [{\it Weak Extensionality\/}]
\[\forall_{Q} x \forall_{Q} y (\forall z (z \in x/_{\equiv} \to \exists t (t \in y/_{\equiv} \wedge \, QSim(z,t) \wedge \forall t(t \in  y/_{\equiv} \to\]
\[\to\exists z (z  \in  x/_{\equiv} \wedge \, QSim(t,z)) \to x \equiv y))\]
\end{description}

	This axiom simply says that those quasi-sets that have the `the same quantity of elements of the same sort' are indistinguishable. 

\begin{teorema}
$\forall_{Q} x \forall_{Q} y (Sim(x,y) \wedge qc(x) =_{E} qc(y) \to x \equiv y)$
\end{teorema}

\noindent
{\bf Proof:} (see \cite{Krause-96})\\

	As a corollary, it follows that $x =_{E} y \to x \equiv y$.

\begin{teorema} 
$\forall_{Q} x \forall_{Q} y (\forall z (z \in x \leftrightarrow z \in y) \to x \equiv y)$
\end{teorema}

\noindent
{\bf Proof:} (See \cite{Krause-96})

\begin{teorema}
$x \equiv y \wedge qc([x]) =_{E} qc([y]) \leftrightarrow [x] \equiv [y]$
\end{teorema}

\noindent
{\bf Proof:} (see \cite{Krause-96})

\subsection{Quasi-Functions}

	With respect to the concept of function, we note that functions, as usually conceived, cannot distinguish between its arguments and values if there were $m$-atoms involved. So, a more general concept of a {\it $q$-function} (quasi-function) as a relation which maps indistinguishable objects into indistinguishable objects is introduced:

\begin{definicao}
Let $x$ and $y$ be quasi-sets. Then we say that $f$ is a $q$-{\em function} from $x$ to $y$ if $f$ is such that ($R$ is the predicate for `relation' defined previously):
 $$R(f) \wedge \forall u (u \in x \to \exists v (v \in y \wedge \langle u, v \rangle \in f)) \wedge $$
$$ \forall u \forall u' \forall v \forall v' (\langle u, v \rangle \in f \wedge \langle u', v'
\rangle \in f \wedge u \equiv u' \to v \equiv v')$$

	If $f$ is a $q$-function from $x$ to $y$ and satisfies the additional condition: 
$$\forall u \forall u' \forall v \forall v' (\langle u, v \rangle \in f \wedge \langle u', v' \rangle \in f \wedge v \equiv v' \to u \equiv u')$$
$$ \wedge  qc(Dom(f)) \leq_{E} qc(Range(f))$$

\noindent
then $f$ is a $q$-{\em injection}, and $f$ is a $q$-{\em surjection} if it is a function from $x$ to $y$ such that $$\forall v (v \in y \to \exists u (u \in x \wedge \langle u, v \rangle \in f)) \wedge
qc(Dem(f)) \geq_{E} qc(Range(f)).$$ An $f$ which is both a $q$-injection and a $q$-surjection is said to be a $q$-{\em bijection}. In this case, $qc(Dom(f)) =_{E} qc(Range(f))$.
\end{definicao}

	In the general case there is no criterion to check if two quasi-sets have the same quasi-cardinal or not, since there is no `counting process' if they have $m$-atoms as elements. This means, for instance, that if (say) $x$ has five elements (formally: its quasi-cardinal is 5), then we cannot define a bijection from $5 = \{0, 1, 2, 3, 4\}$ to $x$, since we would not be able to define without ambiguity the images of $f(0) \ldots f(4)$.

	If $A(x,y)$ is a formula in which $x$ and $y$ are free variables, we say that $A(x,y)$ defines a $x-functional)$ condition on the quasi-set $t$ if $\forall w (w \in t \to \exists s A(w,s) \wedge \forall w \forall w' (w \in t \wedge w' \in t \to \forall s \forall s' (A(w,s) \wedge A(w',s') \wedge w \equiv w' \to s \equiv s'))$ (this is abbreviated by $\forall x \exists ! y A(x,y)$). Then, we have:

\begin{description}

\item[Q28] [{\em Replacement\/}]
$$\forall x \exists ! y A(x,y) \to \forall_{Q} u \exists_{Q} v (\forall z (z \in v \to \exists w (w \in u \wedge A(w,z)))$$

\end{description}

	Intuitively, the  replacement schema says that the images of qsets by q-functions are also qsets. It is easy to see that if there are no $m$-atoms involved, that is, if the qsets are {\it sets\/}, then the above axiom is exactly that of ZFC -- Zermelo-Fraenkel with Axiom of Choice -- (or of ZFU -- Zermelo-Fraenkel with Urelemente). 

\begin{definicao}
A {\em strong singleton} of $x$ is a quasi-set  $x'$ which satisfies the following predicate $St$:
$$St(x') \leftrightarrow x' \subseteq [x] \wedge qc(x') =_{E} 1$$.
\end{definicao}

	That is, $x'$ is a subquasi-set of $[x]$ that has just `one element' which is indistinguishable from $x$. 

\begin{teorema}
For all $x$, there exists a strong singleton of $x$.
\end{teorema}

\noindent
{\bf Proof:} (see \cite{Krause-96})

\begin{description}
\item[Q29] [{\em The Axiom of Choice\/}]
\begin{eqnarray*}
\forall_{Q} x (E(x) \wedge \forall y \forall z (y \in x \wedge z \in x \to y \cap z =_{E} \emptyset  \wedge y \neq_{E} \emptyset)   \to \\
 \mbox{}  
\exists_{Q} u \forall y \forall v (y \in x \wedge v \in y \to \exists_{Q} w (w \subseteq [v] \wedge qc(w) =_{E} 1 \wedge w \cap y \equiv w \cap u)))
\end{eqnarray*}

\end{description}

\begin{teorema}
The extensional equality has all the properties of the usual equality.
\end{teorema}

\noindent
{\bf Proof:} (see \cite{Krause-96})\\

\begin{teorema}{\it [Unobservability of Permutations]} 
Let $x$ be a qset and $z$ an $m$-atom such that $z \in x$. If $w \equiv z$, then $$(x - z') \cup w' \equiv x$$
\end{teorema}

	The operation of difference between qsets is defined as in standard set-theories. This theorem is an immediate consequence of {\bf Q27}. 

	We recall that $z'$ (respect. $w'$) denotes the strong singleton of $z$ (respect., of $w$). Furthermore, it may be the case that $w \notin x$, and this motivates the interpretation according to which the theorem is saying that we have `exchanged' an element of $x$ by an indistinguishable one, and the resulting fact is that `nothing has occurred at all'. In other words, the resulting qset is indistinguishable from the original one. This theorem is the quasi-set theoretical version of the quantum mechanical fact which  expresses that permutations of indistinguishable particles are not regarded as observable, as expressed by the so called Indistinguishability Postulate in quantum mechanics.

\section{Maxwell-Boltzmann Statistics}

\subsection{Some Standard Results in ZF}

	It is a well known theorem in Zermelo-Fraenkel set theory the following:

\begin{lema}
If $x$ is a finite ZF-set, then $$\# {\cal P}(x) = 2^{\# x},$$ where $\# x$ denotes the cardinal of the set $x$.\label{card2}
\end{lema}

\begin{teorema}
Let $x$ be a non-empty and finite ZF-set. If we define $x_2$ as a set of ordered pairs $(y_1,y_2)$ such that $y_1,y_2\in {\cal P}(x)$, $y_1\cup y_2 = x$, and $y_1\cap y_2 = \emptyset$ then $\# x_2 = 2^{\# x}$.\label{yyy}
\end{teorema}

\noindent
{\bf Proof:} Straightforward from Lemma (\ref{card2}).

\begin{teorema}
Let $x$ be a finite ZF-set such that $\# x = N$. If we define $x_n$ as a set of ordered $n$-tuples $(y_1,\cdots, y_n)$ such that for all $i = 1,\cdots, n$ we have $y_i\in{\cal P}(x)$, $\bigcup_i y_i = x$, and $i\neq j\to y_i\cap y_j = \emptyset$, then $\# x_n = n^N$.\label{nN}
\end{teorema}

\noindent
{\bf Proof:} It is straightforward from combinatorial in ZF set theory.

	We could rewrite theorem (\ref{nN}) as:

\begin{teorema}
Let $x$ be a finite ZF-set such that $\# x = N$. If we define $x_n$ as a set of ordered $n$-tuples $(y_1,\cdots, y_n)$ such that for all $i = 1,\cdots, n$ we have $y_i\in{\cal P}(x)$, $\bigcup_i y_i = x$, and $\sum_i \# y_i = \# x$, then $\# x_n = n^N$.\label{nNp}
\end{teorema}

\noindent
{\bf Proof:} Analogous to the proof of theorem (\ref{nN}), since $\bigcup_i y_i = x$, and $i\neq j\to y_i\cap y_j = \emptyset$ iff $\bigcup_i y_i = x$, and $\sum_i \# y_i = \# x$

\subsection{Our Proposal}

	We propose to replace axiom {\bf Q26} in quasi-set theory ${\cal Q}$ by the following assumption (which is a generalization of {\bf Q26} as well as a quasi-set theoretical version of theorem (\ref{nN})):

\begin{description}

\item[Q26'] Let $x$ be a finite quasi-set such that $qc(x) = N$. If we define $z_n$ as the quasi-set whose elements are ordered $n$-tuples $\langle y_1,\cdots , y_n\rangle$, where, for all $i = 1,\cdots, n$, we have $y_i\in {\cal P}(x)$, $\bigcup_{i }y_i = x$, and $\sum_i qc(y_i) = qc(x)$, then we have the following:

\begin{equation}
qc(z_n) = n^N.\label{MBgeneral}
\end{equation}

\end{description}

	In the case where $n = 2$, we have a sentence which is equivalent to axiom {\bf Q26}.

	The main role of axiom {\bf Q26'} is to allow us a quasi-set theoretical combinatorics which can be useful to cope with distribution functions. From the mathematical point of view, it is important to show that the replacement of axiom {\bf Q26} by axiom {\bf Q26'} does not entail any inconsistency in quasi-set theory. This is proved in the Section 5. The point, at this moment, is that {\bf Q26} is very `poor' if we are interested on a quasi-set-theoretical combinatorics with more than two physical states or `boxes', as exemplified in the Introduction. Besides, axiom {\bf Q26'} is our quasi-set theoretical version of theorem (\ref{nNp}).

	If we recall the polynomial of Leibniz, we can rewrite equation (\ref{MBgeneral}) as:

\begin{equation}
qc(z_n) = n^N = \sum\frac{N!}{\Pi_{i = 1,\cdots n}n_i!},\label{MBgeneral2}
\end{equation}

\noindent
where the sum is over all possible combinations of nonnegative integers $n_i$ such that $\sum_{i = 1,\cdots, n}n_i = N$.

	If we interpret $n$ as the number of physical states, $N$ as the total number of particles and $n_i$ as the number of particles associated to each physical state $i$, then it is easy to see that each parcel of the summation in equation (\ref{MBgeneral2}) is a possible MB distribution of $N$ particles among $n$ states. The most probable among all these parcels is the MB distribution. So, we can add equation (\ref{MBgeneral2}), with its respective interpretation, as another extra-assumption in quasi-set theory. In other words, we are generalizing theory ${\cal Q}$, by replacing axiom {\bf Q26} by axiom {\bf Q26'}. We refer to this generalized quasi-set theory as ${\cal Q'}$.

	It is easy to see that, for all $i$ we have $n_i = qc(y_i)$. Axiom {\bf Q26'} is just another manner to say that the number of ways we can distribute $N$ objects (distinguishable or not) among $n$ boxes is $n^N$. The condition that $\bigcup_{i }y_i = x$, and $\sum_i qc(y_i) = qc(x)$ is simply a manner to guarantee that there will be no `repeated occurence' of the same object in two boxes. Nevertheless, it is obvious that the expression `repeated occurence', in this quasi-set-theoretical context, is just an intuitive approach for didatical purposes, since there is no sense in saying that the `same' object cannot occupy two boxes.

	The reader could ask: what are the so-called boxes? Each $y_i$ corresponds to a given box or physical state. There can be, of course, two indistinguishable boxes $y_i$ and $y_j$. In this case, the labels $i$ and $j$ cannot individualize each box. They are just different names, or labels, attributed to two indistinguishable objects (qsets, in this case).

\subsection{One Simple Example}

	Now, let us exhibit an example in order to illustrate our ideas. Consider a collection of three indistinguishable particles to be distributed between two possible states or `boxes'. According to standard textbooks on statistical mechanics there are only four possibilities of distribution. On the other hand, according to our axiomatic framework -- axiom {\bf Q26'} -- there are eight possibilities. If we impose that the occupation number of each box is constant, the number of possibilities corresponds to one parcel of the sum in equation (\ref{MBgeneral2}).

	The question now is: what about the extra four possibilities predicted by axiom {\bf Q26'}? The eight possibilities predicted by {\bf Q26'} and equation (\ref{MBgeneral2}) come from 

$$2^3 = \frac{3!}{3!0!} + \frac{3!}{2!1!} + \frac{3!}{1!2!} + \frac{3!}{0!3!}.$$

	So, we have one possibility with 3 particles in the first state and no particle in the second state, plus three {\em indistinguishable\/}  possibilities with 2 particles in the first state and 1 particle in the second state, plus three {\em indistinguishable\/} possibilities with 1 particle in the first state and 2 particles in the second state, plus one single possibility with no particle in the first state and 3 particles in the remaining one. The calculation of the most probable case is made for a large number of particles, following the standard calculations of statistical mechanics.

	Following our example, axiom {\bf Q26'} says that we can distribute $3$ objects (indistinguishable or not) among $2$ boxes in $2^3$ manners (indistinguishable or not). But this axiom does not say {\em how\/} can we make this distribution. If we do not appeal to equation (\ref{MBgeneral2}), we have the following: according to Fig. 1, there are, at least, by means of axiom {\bf Q16}, {\em four\/} possible distributions. But axiom {\bf Q26'} says that there are eight possible distributions. One possibility is something like Fig. 2, that is, the four distributions in Fig. 1 {\em plus\/} four distributions which are indistinguishable from the third distribution of Fig. 1. The reader can easily imagine other possibilities. So, axiom {\bf Q26'} by itself does not allow us to derive MB statistics. It simply says that MB statistics is a possibility even in a collection of indiscernibles. Axiom {\bf Q26'} {\em and\/} equation (\ref{MBgeneral2}), with its respective interpretation in the context of {\bf Q26'}, is a manner to say that the only possibility is that one illustrated in Fig. 3.

\section{Quantum Statistics}

	What is the difference between quantum statistics and MB, after all? In Bose-Einstein we take into account only distinguishable possibilities, among all possibilities predicted by axiom {\bf Q26'}. And Fermi-Dirac is derived in the same manner, but with the extra assumption of Exclusion Principle in its quasi-set-theoretical form: $qc(y_i)\leq 1$ for each $i$ in {\bf Q26'}. In \cite{Krause-99} the usual quantum distribution functions (Bose-Einstein and Fermi-Dirac) are achieved by means of a quasi-set-theoretical framework. In this paragraph we show that this is not necessary. Put it in another way, quantum statistics may be seen as special cases of MB statistics.

\section{Consistency of Theory ${\cal Q'}$}

\begin{teorema}
${\cal Q'}$ is consistent iff ZFC is consistent.
\end{teorema}

\noindent
{\bf Proof:} Here we make just a very brief sketch of the proof, which can be made in details by the reader, with no difficulty at all. The translation from the language of ZFU to the language of ${\cal Q}$ (as well as to the language ${\cal Q'}$) has shown that if ${\cal Q}$ (and ${\cal Q'}$) is consistent, so is ZFU (and, hence, so is ZFC). In \cite{Krause-96} the converse result for ${\cal Q}$ is outlined. A superstructure $Q$ over a given ZF-set is defined, and a proof that $Q$ is a model for quasi-set theory ${\cal Q}$ is presented. Since our only modification was the replacement of axiom {\bf Q26} by axiom {\bf Q26'}, we concentrate our attention to {\bf Q26'}. The proof of axiom {\bf Q26} in the context of the set-theoretical model $Q$ of ${\cal Q}$ was made by means of a translation of {\bf Q26} into the language of ZFC. Since such a translation simply states a basic property of cardinals in ZFC -- theorem (\ref{yyy}) -- its proof does not represent any problem. In the case of axiom {\bf Q26'} we can use the same argument, since its translation into the language of ZFC (as it was made in \cite{Krause-96}) simply states a basic property of cardinals in ZFC -- theorem (\ref{nNp}).

\section{Final Remarks}

	Our main conclusions are:

\begin{enumerate}

\item By using quasi-set theory instead of standard set theory, our paper provides a way of obtaining the usual statistics in physics from the assumption that the `non-individuality' of quantum objects should be ascribed right at the start.

\item Maxwell-Boltzmann statistics can be derived even in a collection of indiscernibles.

\item Maxwell-Boltzmann statistics may be seen as a generalization of quantum statistics (BE and FD); or, BE and FD are particular cases of MB.

\item Quasi-set theory ${\cal Q'}$ is much more `powerful' than ${\cal Q}$ if we are interested on a quasi-set-theoretical combinatorics.

\item ${\cal Q'}$ is consistent if and only if ZFC is consistent.

\end{enumerate}

	There have been some recent experiments which have demonstrated entangled pairs of atoms \cite{Bouwmeester-97}, which entail the indistinguishability between these atoms. We wonder if it is possible to demonstrate a gas of indistinguishable atoms which preserves the Maxwell-Boltzmann distribution, since it seems that there is no clearly defined frontier between classical and quantum physics.

\section*{Acknowledgments}

	We acknowledge with thanks the important suggestions made by Vilma A. S. Sant'Anna. AMSS acknowledges the financial support from CAPES (Brazilian Government Support Agency).

\newpage

\begin{figure}
\renewcommand{\arraystretch}{0.7}
\[
\begin{array}{|c|c|}\hline
\bullet\bullet\bullet & \;\; \\ \hline
\bullet\bullet & \bullet \\ \hline
\bullet & \bullet\bullet \\ \hline
\;\; & \bullet\bullet\bullet \\ \hline
\end{array}
\]
\caption{The `first' four possible distributions of 3 objects (indistinguishable or not) among 2 boxes. Each line represents one possible distribution and each bullet represents an object.}
\end{figure}
\begin{figure}
\renewcommand{\arraystretch}{0.7}
\[
\begin{array}{|c|c|}\hline
\bullet\bullet\bullet & \;\; \\ \hline
\bullet\bullet & \bullet \\ \hline
\bullet & \bullet\bullet \\ \hline
\;\; & \bullet\bullet\bullet \\ \hline
\bullet & \bullet\bullet \\ \hline
\bullet & \bullet\bullet \\ \hline
\bullet & \bullet\bullet \\ \hline
\bullet & \bullet\bullet \\ \hline
\end{array}
\]
\caption{One possible sequence of the eight possible distributions of 3 objects among 2 boxes according to axiom {\bf Q26'}.}
\end{figure}
\begin{figure}
\renewcommand{\arraystretch}{0.7}
\[
\begin{array}{|c|c|}\hline
\bullet\bullet\bullet & \;\; \\ \hline
\bullet\bullet & \bullet \\ \hline
\bullet\bullet & \bullet \\ \hline
\bullet\bullet & \bullet \\ \hline
\bullet & \bullet\bullet \\ \hline
\bullet & \bullet\bullet \\ \hline
\bullet & \bullet\bullet \\ \hline
\;\; & \bullet\bullet\bullet \\ \hline
\end{array}
\]
\caption{The only possible distribution of 3 objects among 2 boxes, if we conjugate axiom {\bf Q26'} and equation (2).}
\end{figure}


\begin{thebibliography}{99}

\bibitem{Bouwmeester-97} Bouwmeester, D. and Zeilinger, A., `Quantum mechanics: atoms that agree to differ' {\em Nature\/} {\bf 388} 827 - 829 (1997).

\bibitem{Krause-92} Krause, D., `On a  quasi-set theory', {\em Notre Dame Journal of Formal Logic\/} {\bf 33} 402-411 (1992). 

\bibitem{Krause-96} Krause, D., `Axioms for collections of indistinguishable objects', {\em Logique et Analyse\/} {\bf 153--154}, 69-93 (1996).

\bibitem{Krause-99} Krause, D., A. S. Sant'Anna and A. G. Volkov, `Quasi-set theory for bosons and fermions: quantum distributions', {\em Found. Phys. Lett.\/}, {\bf 12} 67-79 (1999).

\bibitem{Mendelson-97} Mendelson, E., {\em Introduction to Mathematical Logic\/} (Chapman \& Hall, London, 1997).

\bibitem{Sant'Anna-97a} Sant'Anna, A. S., `Some remarks about indistinguishability and elementary particles', {\em Logique et Analyse} {\bf 157} 45-66 (1997).
\bibitem{Sant'Anna-97b} Sant'Anna, A. S. and D. Krause, `Indistinguishable particles and hidden variables', {\em Found. Phys. Lett.} {\bf 10} 409-426 (1997).

\end{thebibliography}
\end{document}